# Vortex Trapping in Hybrid Magnetometers

K.H. Kuit[1*], J.R. Kirtley[2], H. Rogalla[1], J. Flokstra[1]

[1] Low Temperature Division, MESA+ Institute for Nanotechnology, University of Twente, P.O. Box 217, 7500AE Enschede, The Netherlands

[2] Department of Applied Physics, Stanford University, Palo Alto, California 94305, USA

Corresponding author :    K.H. Kuit,

Phone : + (31) 53 4894627

Fax : +(31) 53 4891099

e-mail : k.h.kuit@tnw.utwente.nl

**Abstract**

Hybrid magnetometers based on a normal conducting sensor and a superconducting flux concentrator have been investigated. When this sensor is operated in an unshielded environment flux vortices can be trapped in the superconducting body when the sensor is cooled. Thermal hopping of the trapped vortices gives rise to 1/f noise. The mechanism for vortex trapping has been investigated. A new model for the critical field for vortex trapping and vortex densities has been derived. Experimental verification on $YBa_2Cu_3O_{7-\delta}$ (YBCO) strips with a Scanning SQUID Microscope shows very good agreement between measurements and the new theory. The results have been applied in a hybrid magnetometer design based on a bismuth Hall sensor and an YBCO flux concentrator. Measurements show a field gain up to $G \approx 7$ with a large effective range of $\pm 175 \mu T$.

# 1- Introduction

A wide variety of magnetometers is available varying in sensitivity, size and power consumption. We investigate the possibilities of a magnetic field sensor suitable for operation in micro-satellites. This means that a small, robust and low power consuming sensor is required which nevertheless has a high sensitivity. It is hard to find a sensor which has all these properties combined. Magneto resistances like for example a GMR or AMR and Hall sensor are robust sensors but their sensitivity can not compete with for example SQUIDs and optically pumped magnetometers. Nevertheless these relatively insensitive sensors can be interesting if the sensitivity can be improved by flux concentration [1,2].

The principle of flux concentration is relatively simple. The flux concentrator is a superconducting ring. When a field is applied to the ring a screening current will screen the field from the ring. In the ring a narrow constriction is constructed through which all the current is forced. At this specific point the magnetic field generated by the screening current is higher than the background field and this is where a sensor is placed.

When this sensor is cooled in an unshielded environment flux vortices can get trapped in the superconducting body. Due to thermal hopping of the vortices the 1/f noise of the sensor increases. A known method to prevent vortex trapping is to make slots in the superconducting body [3,4]. A thorough investigation of vortex trapping in thin film superconducting strips has been conducted in order to come to a maximum magnetic field during cooling without a decrease in low frequency sensitivity. In literature two models are known for the critical field which do not fit the few experimental data very well [5]. A new model has been derived which gives the critical field for vortex trapping and the vortex density above the critical field. To verify the model, measurements have been performed with a Scanning SQUID Microscope (SSM) and excellent agreement between model and measurements was found.

The results from this part of the research have been used in the design of a hybrid magnetometer. Several designs have been made with and without slots in the washer of the concentrator and in varying widths. Measurements showed that the magnetometers perform as expected. The slots in the washer however give an unwanted effect at the ends of the effective field range.

## 2- Vortex trapping

### 2.1 Theory

The flux concentrators used in this research are made by structuring thin films of high temperature superconductors. Whether or not vortices get trapped in the resulting narrow thin film strip is determined by the Gibbs free energy. The Gibbs free energy of a strip has a dome shape for low applied fields. When the field is increased at a certain value a dip will start to appear in the middle of the strip which can act as an energy barrier for the escape of vortices. There are two known theories in literature which give a value for the critical field for the escape of vortices. The first model is a metastable condition where it is assumed that there should be no dip at all in the energy [6]. The second proposed model assumes an absolute stable condition and states that the Gibbs free energy in the middle of the strip should equal zero [7].

The model proposed here is an intermediate model between Ref [6] and [7]. When a strip is cooled to just below $T_c$, due to thermal fluctuations there will be generation of vortex-antivortex pairs. The antivortices are quickly driven out of the strips but whether or not the vortices can escape, in order to recombine, depends on the height of the energy barrier. In the model we propose the energy required to form a vortex-antivortex pair should equal the height of the energy barrier in the Gibbs free energy, resulting in a critical field [8]:

$$B_K = 1.65 \frac{\Phi_0}{W^2}. \qquad (1)$$

From the Gibbs free energy also the vortex density above the critical field can directly be derived with a $1/\Phi_0$ dependency [8]:

$$n = \frac{B_a - B_K}{\Phi_0} \qquad (2)$$

### 2.1 Experiment

To evaluate the models from literature and the presented model measurements in a SSM have been performed. A sample has been produced with a pulsed laser deposited YBCO film of approximately 200 nm on a SrTiO$_3$ (STO) substrate structured into narrow strips varying from 2-50µm with Ar ion etching. This sample is cooled down in a magnetic field and is scanned by the SSM. The vortices are pinned to their final positions in the strips below the depinning temperature, which is just below the critical temperature, however the sample is further cooled down to 4.2 K for the measurement. In figure 1-a) a graph is displayed of the critical field of the strips together with the models. For every strip width there are 2 points in the plot. The upper data point indicates the lowest possible field with some trapped vortices visible and the lower data point the highest possible field without any trapped vortices. It is clear from this graph that the model presented in equation 1 shows the best fit with the data. For 6 and 35 µm wide strips the vortex density as a function of the applied field is plotted in figure 1-b). In this case the measurements show a reasonable fit with the model from equation 2. The vortex-vortex interaction is not included in this model which might improve the fit.

## 3- Hybrid Magnetometer devices

Several magnetometer designs have been produced with slotted and solid washers. The concentrator ring is made on an STO substrate with a pulsed laser deposited layer of YBCO covered by gold which is structured by Ar ion etching. As an insulator between the concentrator and the sensor a layer $SiO_2$ is deposited and structured by lift off. Next to the constriction a Hall element of Bi is deposited by rf-sputtering and structured by lift off. In figure 2-a) a Scanning Electron Microscope (SEM) image of a typical device is displayed. The concentrator has a diameter of 7x7 mm and in this particular case the washer is slotted into 35 µm wide strips. The sensor is situated in the bottom part of the picture with wires and bonding pads for the 2 current and 1 voltage lines. The second voltage connection is placed in the middle of the concentrator. In figure 2-b) a zoom of the part where the sensor is located is displayed (90 degrees rotated). The $SiO_2$ layer is deposited over the constriction so the bismuth layer can make a connection to the strip line on the other side. The Au on top of the YBCO layer is solely their to prevent direct connection between the Bi and the YBCO layer, so no diffusion of oxygen from the YBCO to the Bi can occur.

### 3.1 Measurements

To characterize the samples the Hall voltage is measured during sweeps of the field. The results of a typical measurement are displayed in figure 3. In this graph the sensor is measured for 3 bias currents through the Hall element. Basically the sensor is used beyond the maximum field range so the behavior of the device with and without the flux concentrator can be compared. In the steep part of the characteristic the sensor is working with the flux concentrator. At a certain applied field the screening current in the ring becomes so large that the constriction reaches its critical current so field lines can penetrate the hole of the ring and the screening current is lowered to just below the critical current. Any further increase of the field will not increase screening current any more and more flux lines enter the hole of the ring. Above this critical field only the Hall effect of the sensor is visible which is obviously less steep. From this kind of measurements the gain of the ring, the sensitivity, the Hall coefficient and the effective range of operation can be retrieved. The gain of the flux concentrator on the sensor is ~7 with an effective range of operation of ±175µT. The gain and the effective range are independent of the bias current through the sensor which means that there are no heating effects of the dissipation in the sensor on the constriction. During the design process the effective range was aimed to be ±100µT. This demand meant that the constriction should be 70 µm wide to reach the desired critical current. However the critical current density of YBCO turns out to be much higher so a narrower constriction will be sufficient to come to the same effective range of operation and this increases the gain probably up to ~30. The sensitivity of the sensor is 90mV/T when the bias current through the Hall sensor is 10 mA. The poor sensitivity is for a part caused by the low gain but also by the low Hall coefficient of ~$7 \times 10^{-7}$ m$^3$/C which can be improved by optimization of the deposition conditions of Bi. In between the two regions of operation a third intermediate steep slope can be observed. This region is believed to be caused by the outer strips in the concentrator which reach their critical current at a certain field due to inhomogeneous current distribution in the ring. This affects the effective area and the induction of the concentrator making it less sensitive for field. This effect is under investigation.

## 4- Conclusions

A new model for the critical field of vortex trapping and vortex densities is presented. Experimental verification with SSM measurements shows very good agreement with the model. The results are incorporated in hybrid magnetometer designs. The produced devices show that the principle of flux concentration is working; however there is room for improvement. New designs with narrower constriction should give a higher field gain and optimization of the deposition conditions of Bi can improve the Hall coefficient of the sensor.

**Acknowledgments**

This research was financed by the Dutch MicroNED program project 1-D-2.

**Legends**

Figure 1 : a) Critical inductions for vortex trapping as a function of strip width. The squares represent $B_{c+}$, the lowest inductions in which trapped vortices were observed, and the dots are $B_{c-}$, the highest inductions in which trapped vortices were not observed. The dash-dotted line is the metastable critical induction [6]. The short-dashed and long-dashed lines are the absolute stable critical inductions [7] for two assumptions of the vortex core size. The solid line is from our model [8]. b) Vortex density as a function of the applied field for 6 and 35 µm wide strips. The solid line is the prediction of equation 2.

Figure 2 : SEM images of a typical example of a hybrid magnetometer. a) Overview of the device with the sensor located in bottom part of the image. b) Zoom of the sensor (90 degrees rotated with respect to figure a) ) .

Figure 3 : Hall voltage as a function of the applied field for 5, 7.5 and 10 mA bias current through the sensor. The device is operated beyond the effective operation range resulting in the hysteretic type of behavior.

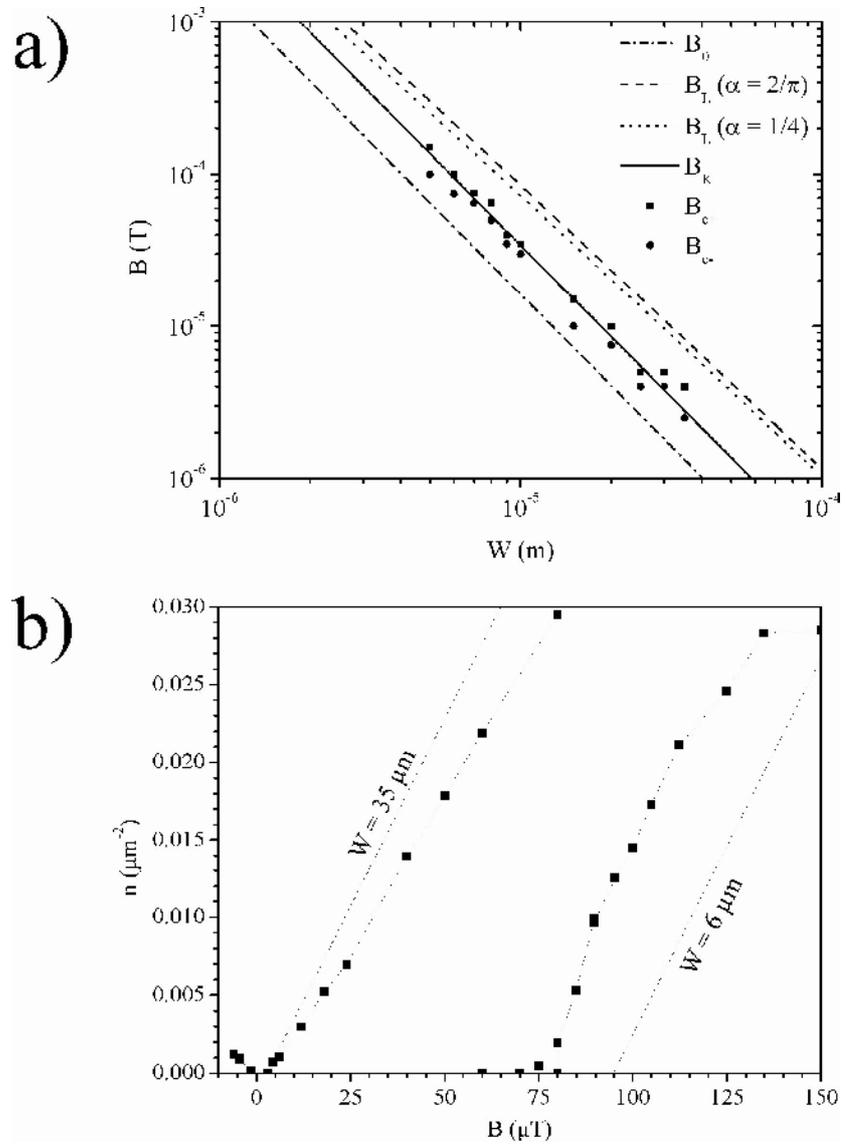

Figure 1

Kuit et al.

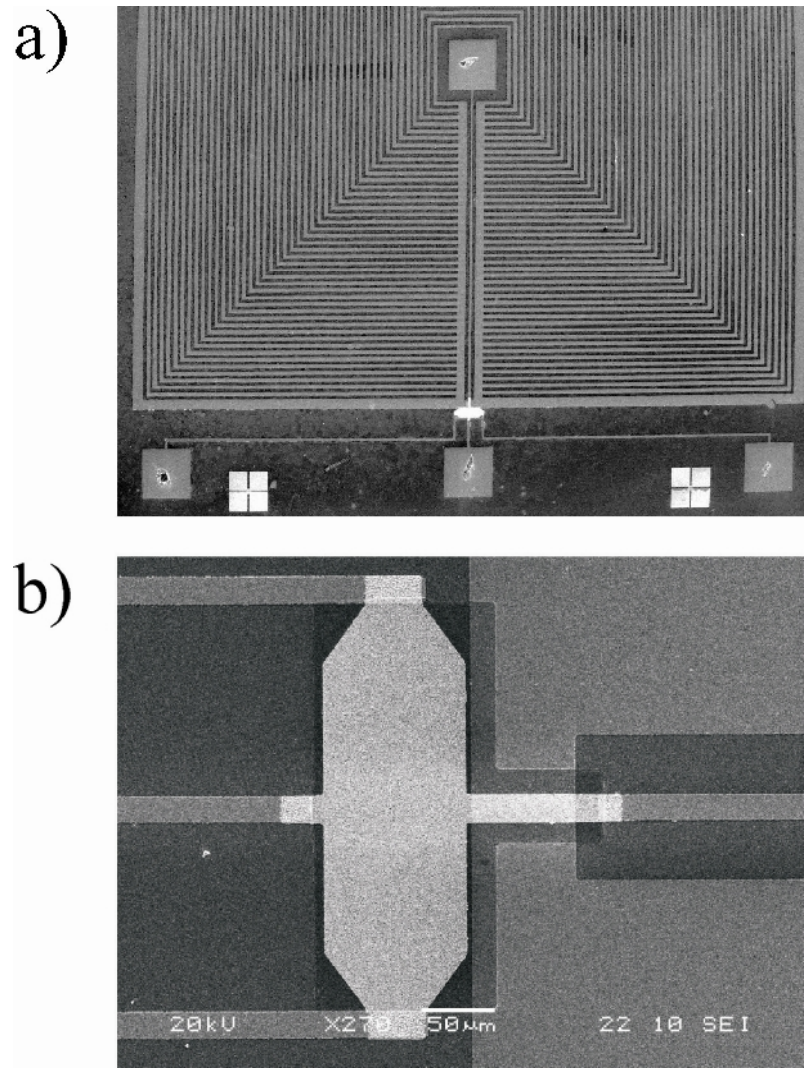

Figure 2
Kuit et al.

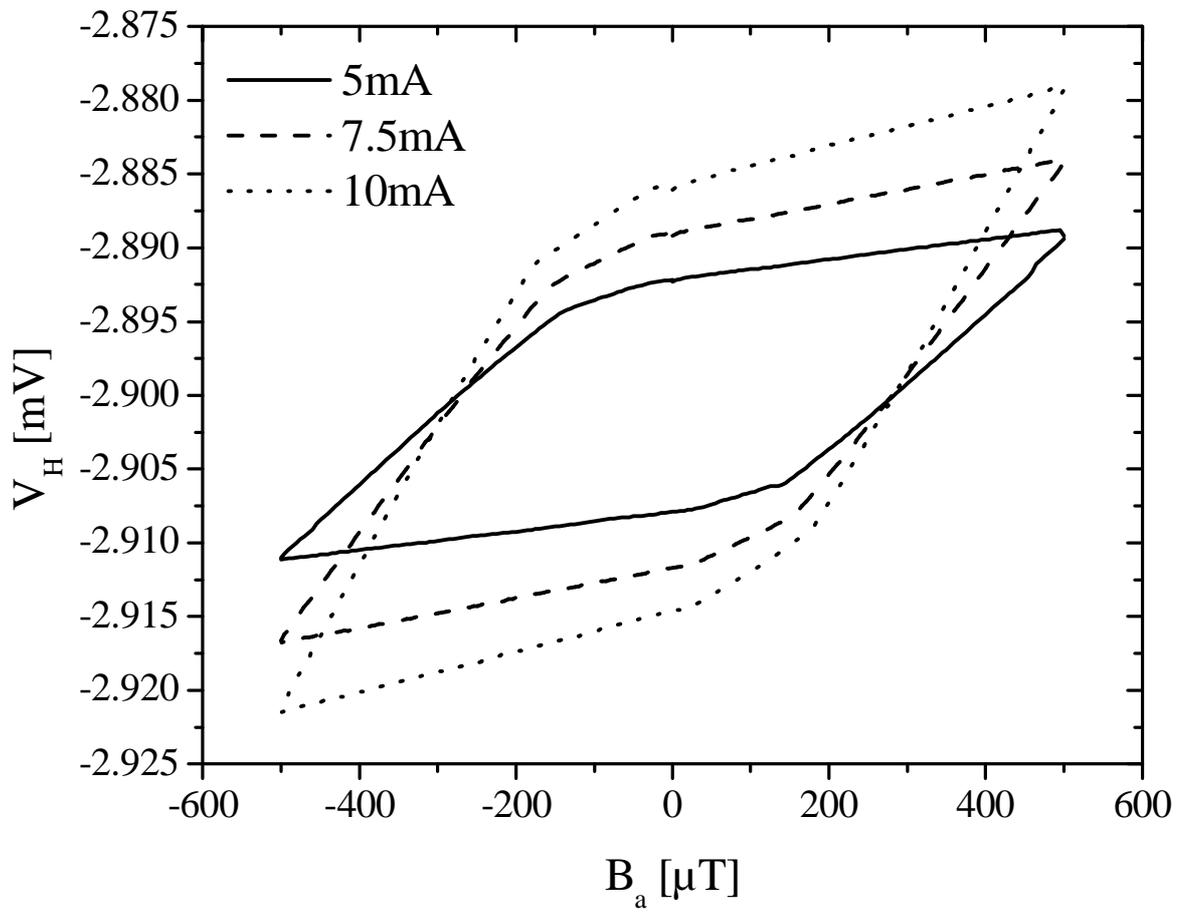

Figure 3

Kuit et al.